\documentclass[twocolumn,secnumarabic,amssymb, amsmath, nofootinbib,tightenlines,
nobibnotes, aps, prl,epsfig]{revtex4}
\usepackage{graphicx}% Include figure files
\usepackage{dcolumn}% Align table columns on decimal point
\usepackage{bm}% bold math
\begin{document}
\preprint{APS/123-QED}
\title{A Phenomenological Solution Small $x$ to the Longitudinal Structure Function  Dynamical Behaviour  }% Force line breaks with \\

\author{G. R. Boroun }
\altaffiliation{boroun@razi.ac.ir}%Lines break automatically or can be forced with \\
\author{B. Rezaei}%
\affiliation{ Physics Department, Razi University, Kermanshah
67149, Iran}% \textbackslash\textbackslash
\author{J. K. Sarma}%
\affiliation{ HEP Laboratory, Department of Physics, Tezpur University,  Napaam 784 028, Assam, India}% \textbackslash\textbackslash
\date{\today}% It is always \today, today,

\begin{abstract}
In this paper, the evolutions of longitudinal proton structure
function have been obtained at small-$x$ upto
next-to-next-to-leading order using a hard pomeron behaviour. In
our paper, evolutions of gluonic as well as heavy longitudinal
structure functions have been obtained separately and the total
contributions have been calculated. The total longitudinal
structure functions have been compared with results of
Donnachie-Landshoff (DL) model, Color Dipole (CD) model, $k_{T}$
factorization and H1 data.

\ PACS number(s): 12.38.-t, 12.38.Bx, 14.70.Dj

\end{abstract}
% \pacs{11.55Jy, 12.38.-t, 14.70.Dj}%PACS, the Physics and Astronomy
                              %Classification Scheme.

\maketitle
%%%%%%%%%%%%%%%%%%%%%%%%%%%%%%%%%%%%%%%%%%%%%%%%%%%%%%%%%%%%%%%%%
\subsection{I. Introduction}
The measurement of the longitudinal structure function
$F_{L}(x,Q^{2})$ is of great theoretical importance, since it may
allow us to distinguish between different models describing the
QCD evolution  small $x$. In deep-inelastic scattering (DIS), the
structure function measurements remain incomplete until the
longitudinal structure function $F_{L}$ is actually measured [1].
The longitudinal structure function in DIS is one of the
observables from which the gluon distribution can be unfolded. The
dominant contribution small $x$ to $F_{L}(x,Q^{2})$ comes from the
gluon operators. Hence a measurement of $F_{L}(x,Q^{2})$ can be
used to extract the gluon structure function and therefore the
measurement of $F_{L}$ provides a
sensitive test of perturbative QCD (pQCD) [2-3].\\
The experimental determination of $F_{L}$ is in general difficult
and requires a measurement of the inelastic cross section at the
same values of $x$ and $Q^{2}$ but for different center-of-mass
energy of the incoming beams. This was achieved at the DESY
electron-proton collider HERA by changing the proton beam energy
with the lepton beam energy fixed. HERA collected $e^{+}p$
collision data with  the H1 and ZEUS detectors at a positron beam
energy of $27.5 GeV$ and a proton beam energies of $920, 575$ and
$460 GeV$, which allowed a measurement of structure functions at
$x$ values $5{\times}10^{-6}{\leq}x{\leq}0.02$ and $Q^{2}$ values
$0.2~ GeV^{2} {\leq}Q^{2}{\leq}800 ~GeV^{2}$ [4].\\
Since the longitudinal structure function $F_{L}$ contains rather
large heavy flavor contributions in the small-$x$ region,
therefore the measurement of these observables have told us about
the different scheme used to calculate the heavy quark
contribution to the structure function and also the dependence of
parton distribution functions (PDFs) on heavy quark masses [5].
For PDFs we need to use the corresponding massless Wilson
coefficients
 up to next-to-next-to leading order (NNLO) [6-14],
but we determine heavy contributions of longitudinal structure
function in leading order and next-to-leading order by using
massive Wilson coefficients in the asymptotic region
$Q^{2}{\gg}m_{h}^{2}$, where $m_h$ is the mass of heavy quark [15-21].\\
The dominant small $x$ role is played by gluons, and the basic
dynamic quantity is the unintegrated gluon distribution
$f(x,Q_{t}^{2})$, where $Q_{t}$ its transverse momentum. In the
leading $\ln(1/x)$ approximation, the Lipatov equation, which
takes into account all the $LL(1/x)$ terms has the following form
[22-27]
\begin{eqnarray}
f(x,Q_{t}^{2})&=&f^{0}(x,Q_{t}^{2})+\frac{N_{c}\alpha_{s}}{\pi}\int_{x}^{1}\frac{dx'}{x'}\int\frac{d^{2}q}{{\pi}q^{2}}[\frac{Q_{t}^{2}}{(\mathbf{q}+\mathbf{Q_{t}})^{2}}\nonumber\\
&&f(x',(\mathbf{q}+\mathbf{Q_{t}})^{2})-f(x',Q_{t}^{2})\Theta(Q_{t}^{2}-q^{2})],\nonumber\\
\end{eqnarray}
where
\begin{equation}
f(x,Q_{t}^{2}){\equiv}
\frac{{\partial}[xg(x,Q^{2})]}{{\partial}\ln
Q^{2}}|_{Q^{2}=Q_{t}^{2}}.
\end{equation}
The NLO corrections can be find in [28-29]. This equation sums the
ladder diagrams with a gluon exchange accompanied by virtual
corrections that are responsible for the gluon reggeization. The
analytical solution  small $x$ is given by
\begin{eqnarray}
f(x,Q_{t}^{2}){\sim}\mathcal{R}(x,Q_{t}^{2})x^{-\lambda_{BFKL}}
\end{eqnarray}
where $\lambda_{BFKL}=4\frac{N_{c}\alpha_{s}}{\pi}\ln(2)$ at LO
and at NLO it has the following form
\begin{equation}
\lambda_{BFKL}=4\frac{N_{c}\alpha_{s}}{\pi}\ln(2)[1-\frac{\alpha_{s}\beta_{0}}{\pi}(\ln2-\frac{\pi^{2}}{16})].
\end{equation}
The quantity $1+\lambda_{BFKL}$ is equal to the intercept of the
so-called BFKL Pomeron.  In the phenomenological analysis of the
high energy behavior of hadronic as well as photoproduction total
cross section, the value of the intercept is determined as
 $\alpha_{soft}{\approx}1.08$ (as this is  the effective
soft Pomeron) [30]. In DIS, a second hard-Pomeron must be added
with a larger intercept $\alpha_{hp}{\approx}1.4$ [31-35] and
recently decreases to the value $1.317$ where estimated directly
from the data on the proton unpolrized structure function [36].\\
It is tempting, however, to explore the possibility of obtaining
approximate analytic solutions of the
Dokshitzer-Gribov-Lipatov-Altarelli-Parisi (DGLAP) [37-39]
equations themselves in the restricted domain of small-$x$ at
least. Approximate solutions of the DGLAP equations have been
reported
[40-48] with considerable phenomenological success.\\
The objective of this paper is the calculation of the evolution
equation of $F_{L}$ using a hard Pomeron behavior  small $x$ at LO
up to NNLO.  Therefore we concentrate on the Pomeron in our
calculations, although clearly good fits relative to results show
that the gluon distribution and the singlet structure function
need a model having hard Pomeron. Our paper is organized as
follows : in section $I$, which is the introduction, we described
the background in short. Section $II$ is the theory where we have
discussed the non-singlet, light and heavy part of longitudinal
structure function separately in details.
 Section $III$ is the results and discussions on the results. Section $IV$ is the conclusions where overall conclusions were given in brief. Lastly in appendix A and in appendix B we have put the required explicit forms of splitting functions and coefficient functions respectively.\\

\subsection{II. Theory}
In perturbative QCD, the longitudinal structure function can be
written as [6,11-16]
\begin{eqnarray}
x^{-1}F_{L}&=&C_{L,ns}{\otimes}q_{ns}+<e^{2}>(C_{L,q}{\otimes}q_{s}+C_{L,g}{\otimes}g)\nonumber\\
&&+x^{-1}F^{heavy}_{L}
\end{eqnarray}
where ${\otimes}$
denotes the common convolution in the $N_{f}=3$ light quark flavor sector, and $q_{i}$ and $g$ represent the
number distributions of quarks and gluons, respectively, in the
fractional hadron momentum. $q_{s}$ stands for the flavour-singlet
quark distribution, $q_{s}=\sum_{u,d,s}(q+\overline{q})$, and
$q_{ns}$ is the corresponding non-singlet combination. The average
squared charge ($=\frac{2}{9}$ for light quarks) is represented by
$<e^{2}>$. The symbol ${\otimes}$ represents the standard Mellin
convolution and is given by
\begin{eqnarray}
A(x){\otimes}B(x)=\int_{x}^{1}\frac{dy}{y}A(y)B(\frac{x}{y}).
\end{eqnarray}
The perturbative expansion of the coefficient functions can be
written [11-14] as
\begin{eqnarray}
C_{L,a}(\alpha_{s},x)=\sum_{n=1}(\frac{\alpha_{s}}{4\pi})^{n}c^{(n)}_{L,a}(x).
\end{eqnarray}
In Eq.7, the superscript of the coefficients on the right-hand
side represents the order in $\alpha_{s}$ and not, as for the
splitting functions, the `$m$' in $N^{m}LO$ [11-13]. According to
Eq.5 we display the individual parton distributions separately,
then discuss those evolutions as
\begin{equation}
F^{total}_{L}=F^{ns}_{L}+F^{light}_{L}(=F^{q}_{L}+F^{g}_{L})+F^{heavy}_{L}.
\end{equation}

{\bf A)} {\bf Evolution of non-singlet longitudinal structure function:}\\
 The non-singlet longitudinal structure
function $F_{L,ns}$ obtained from the connection between the quark
coefficient function $C_{L,ns}$ and quark distribution $q_{ns}$ is
given by [49],
\begin{eqnarray}
\mathcal{F}^{ns}_{L}(x,Q^{2}){\equiv}~x^{-1}F^{ns}_{L}(x,Q^{2})=C_{L,ns}{\otimes}q_{ns}(x,Q^{2}).
\end{eqnarray}
By differentiating Eq.9 with respect to $Q^{2}$ by means of the
evolution equations for $a_{s}=\frac{\alpha_{s}}{4\pi}$ and
$q_{ns}(x,Q^{2})$,
\begin{eqnarray}
\frac{da_{s}}{d{\ln}Q^{2}}&=&\beta(a_{s})=-\beta_{0}a_{s}^{2}-\beta_{1}a_{s}^{3}-\beta_{2}a_{s}^{4}-..,
\end{eqnarray}

where
\begin{eqnarray}
\beta_{0}&=&\frac{11}{3}N_{c}-\frac{4}{3}T_{f},\nonumber\\
\beta_{1}&=&\frac{34}{3}N_{c}^{2}-\frac{20}{3}N_{c}T_{f}-4C_{F}T_{f},\nonumber\\
\beta_{2}&=&\frac{2857}{54}N_{C}^{3}+2C_{F}^{2}T_{f}-\frac{205}{9}C_{F}N_{C}T_{f}\nonumber\\
&&-\frac{1415}{27}N_{C}^{2}T_{f}+\frac{44}{9}C_{F}T_{f}^{2}+\frac{158}{27}N_{C}T_{f}^{2},
\end{eqnarray}
are the one-loop, two-loop and three-loop corrections to the QCD
$\beta$-function and
\begin{eqnarray}
N_{C}=3,~ ~C_{F}=\frac{N_{C}^{2}-1}{2N_{C}}=\frac{4}{3},~~
T_{f}=T_{R}N_{f}=\frac{1}{2}N_{f},
\end{eqnarray}
where $N_{C}$ is the number of colors and $N_{f}$ is the number of
active flavors,

and
\begin{eqnarray}
\frac{dq_{ns}}{d{\ln}Q^{2}}=P_{ns}{\otimes}q_{ns}.
\end{eqnarray}
The non-singlet evolution equation for the longitudinal structure
function  large $x$ is obtained as

\begin{eqnarray}
\frac{d\mathcal{F}^{ns}_{L}(x,Q^{2})}{d{\ln}Q^{2}}&=&\{P_{ns}(a_{s})+\beta(a_{s})\frac{d}{da_{s}}{\ln}c_{L,ns}(a_{s})\}{\otimes}\mathcal{F}^{ns}_{L}\nonumber\\
&&=K_{L,ns}{\otimes}\mathcal{F}^{ns}_{L}(x,Q^{2}).
\end{eqnarray}\\

{\bf B)} {\bf Evolution of light longitudinal structure function:}\\
Now, we present our evolution for the light longitudinal structure
function in electromagnetic DIS at three loops, where `light'
refers to the common $u,d,s$ quarks and their anti quarks, and
gluon distributions, as [5-16, 49-53, 11-12]
\begin{eqnarray}
F^{light}_{L}(x,Q^{2})&=&C_{L,q}{\otimes}F^{s}_{2}(x,Q^{2})+<e^{2}>C_{L,g}{\otimes}G(x,Q^{2})\nonumber\\
&&=\sum_{n=1}(\frac{\alpha_{s}}{4\pi})^{n}[c^{(n)}_{L,q}(x){\otimes}F^{s}_{2}(x,Q^{2})\nonumber\\
&&+<e^{2}>c^{(n)}_{L,g}(x){\otimes}G(x,Q^{2})]\nonumber\\
&&{\equiv}\sum_{n=1}F^{(n),light}_{L}(x,Q^{2}),
\end{eqnarray}
where $F_{2}^{s}$ refer to the singlet structure function and
$G(=xg)$ is the gluon distribution function. The singlet part of
Wilson coefficients is, that decomposed into the non-singlet and a
pure singlet contribution, denoted by
\begin{eqnarray}
c^{(n)}_{L,q}=c^{(n)}_{L,ns}+c^{(n)}_{L,ps}.
\end{eqnarray}
We start by differentiating Eq.15 with respect to ${\ln}Q^{2}$ as
\begin{widetext}
\begin{eqnarray}
\frac{{\partial}F^{light}_{L}(x,Q^{2})}{{\partial}\ln
Q^{2}}&=&\sum_{n=1}n\frac{{d}{\ln}\alpha_{s}}{{{d}\ln
Q^{2}}}[(\frac{\alpha_{s}}{({4\pi})})^{n}[c^{(n)}_{L,q}(x){\otimes}F^{s}_{2}(x,Q^{2})+<e^{2}>c^{(n)}_{L,g}(x){\otimes}G(x,Q^{2})]]\nonumber\\
&&+\sum_{n=1}(\frac{\alpha_{s}}{4\pi})^{n}
[c^{(n)}_{L,q}(x){\otimes}\frac{{\partial}F^{s}_{2}(x,Q^{2})}{{\partial}\ln
Q^{2}}+<e^{2}>c^{(n)}_{L,g}{\otimes}\frac{{\partial}G(x,Q^{2})}{{\partial}\ln
Q^{2}}]\nonumber\\
&&=\frac{{d}{\ln}\alpha_{s}}{{{d}\ln Q^{2}}}[\sum_{n=1}
n{\times}F^{(n),light}_{L}(x,Q^{2})]+\sum_{n=1}(\frac{\alpha_{s}}{4\pi})^{n}
[c^{(n)}_{L,q}(x){\otimes}\frac{{\partial}F^{s}_{2}(x,Q^{2})}{{\partial}\ln
Q^{2}}+<e^{2}>c^{(n)}_{L,g}{\otimes}\frac{{\partial}G(x,Q^{2})}{{\partial}\ln
Q^{2}}].\nonumber\\
\end{eqnarray}
\end{widetext}
The general mathematical structure of the DGLAP equation is
[37-39]
\begin{eqnarray}
\frac{{\partial}xf(x,Q^{2})}{{\partial}\ln Q^{2}}=
P(x,\alpha_{s}(Q^{2})){\otimes}xf(x,Q^{2}).
\end{eqnarray}
The perturbative expansion of the kernels and of the beta function
at LO up to NNLO are respectively
\begin{eqnarray}
P(x,\alpha_{s})&=&(\frac{\alpha_{s}}{2\pi})P^{(LO)}(x)+(\frac{\alpha_{s}}{2\pi})^{2}P^{(NLO)}(x)\nonumber\\
&&+(\frac{\alpha_{s}}{2\pi})^{3}P^{(NNLO)}(x)+... ~.
\end{eqnarray}
 The DGLAP evolution equations for the singlet
quark structure function and the gluon distribution are given by
[37-39,50-51]
\begin{widetext}
\begin{equation}
\frac{{\partial}G(x,Q^{2})}{{\partial}{\ln}Q^{2}}=P_{gg}(x,\alpha_{s}(Q^{2})){\otimes}
G(x,Q^{2})+P_{gq}(x,\alpha_{s}(Q^{2})){\otimes} F_{2}^{s}(x,Q^{2})
\end{equation}

and

\begin{equation}
\frac{{\partial}F_{2}^{s}(x,Q^{2})}{{\partial}{\ln}Q^{2}}=
P_{qq}(x,\alpha_{s}(Q^{2})){\otimes}
F_{2}^{s}(x,Q^{2})+2n_{f}P_{qg}(x,\alpha_{s}(Q^{2})){\otimes}
G(x,Q^{2}),
\end{equation}
\end{widetext}
After substituting Eqs.20 and 21 in Eq.17 and using Eq.15, we can
not find an evolution equation for the singlet longitudinal
structure function, because it contains both singlet and gluon.
But, at small values of $x$ ($x \leq 10^{-3}$), the gluon
contribution to the light $F_{L}$ structure function dominates
over the singlet and non-singlet contribution [49]. Therefore
$F^{light}_{L} {\rightarrow} F^{g}_{L}$ and we have the gluonic
longitudinal structure function as
\begin{eqnarray}
F^{g}_{L}(x,Q^{2})&=&\sum_{n=1}(\frac{\alpha_{s}}{4\pi})^{n}<e^{2}>c^{(n)}_{L,g}(x){\otimes}G(x,Q^{2})\nonumber\\
&&{\equiv}\sum_{n=1}F^{(n),g}_{L}(x,Q^{2}).
\end{eqnarray}
By differentiating this equation with respect to $Q^{2}$ by means
of the evolution equations for $\alpha_{s}(Q^{2})$ and
$G(x,Q^{2})$ according to Eq.20 at small-$x$ and assuming gluon
distribution is dominant, we find that
\begin{eqnarray}
\frac{{\partial}F^{g}_{L}(x,Q^{2})}{{\partial}\ln
Q^{2}}&=&\frac{{d}{\ln}\alpha_{s}}{{d}\ln
Q^{2}}[\sum_{n=1}n{\times}F^{(n),g}_{L}(x,Q^{2})]\nonumber\\
&&+P_{gg}{\otimes}F^{g}_{L}(x,Q^{2}).
\end{eqnarray}
 The explicit forms of
the splitting functions up to third-order are given in Appendix A.
Eq.23 leads to the gluonic longitudinal evolution equation small
$x$, where it can be calculated by hard-Pomeron behavior for the
gluon distribution function up to NNLO. This issue is the subject
of the next section.\\
\vspace{2cm}

{\bf C)} {\bf Evolution of heavy longitudinal structure function:}\\

One of the important areas of research at accelerators is the
study of heavy flavor production. Heavy flavors can be produced in
electron-positron, hadron-hadron, photon-hadron and lepton-hadron
interactions. We concentrate on the last and in particular on
electron-proton collisions which investigate heavy flavor
production experimentally at HERA. In pQCD calculations the
production of heavy quarks at HERA (and recently at LHC) proceeds
dominantly via the direct boson-gluon fusion (BGF), where the
photon interacts indirectly with a gluon in the proton by the
exchange of a heavy quark pair [54-61]. The data for heavy quark
(c, b) production in the BGF dynamic, have been theoretically
described in the fixed-flavor number factorization scheme by the
fully predictive fixed-order perturbation theory. With respect to
the recent measurements of HERA, the charm contribution to the
structure function  small $x$ is a large fraction of the total, as
this value is approximately $30\%$ ($1\%$) fraction of the total
for the charm (bottom) quarks respectively. This behavior is
directly related to the growth of the gluon distribution at
small-$x$. We know that the gluons couple only through the strong
interaction, consequently the gluons are not directly probed in
DIS. Therefore, the study of charm production in deep inelastic
electron-proton scattering indirectly via the
 $g{\rightarrow}q\bar{q}$
transition is given by the reaction
\begin{eqnarray}
e^{-}(l_{1})+P(p){\rightarrow}e^{-}(l_{2})+C(p_{1})\overline{C}(p_{2})+X,
\end{eqnarray}
where $X$ stands for any final hadronic state.\\
We now derive our master formula for evolution of the $F_{L}^{c}$
for small values of $x$, which has the advantage of being
independent of the gluon distribution function. In the range of
small-$x$ , where only the gluon and quark-singlet contributions
matter, while the non-singlet contributions are negligibly small,
we have [18]
\begin{eqnarray}
F_{L}^{c}(x,Q^{2})=\sum_{a}\sum_{l}C_{L,a}^{l}(x,Q^{2}){\otimes}xf_{a}^{l}(x,Q^{2}),
\end{eqnarray}
with parton label $a=g, q, \overline{q}$, where $q$ generically
denotes the light-quark flavours and $l = \pm $ labels the usual
$+$ and $-$ linear combinations of the gluon and quark-singlet
contributions, $C^{l}_{L,a}(x,Q^{2})$ is the DIS coefficient
function, which can be calculated perturbatively in the parton
model of QCD (Appendix B). A further simplification is obtained by
neglecting the contributions due to incoming light quarks and
antiquarks in Eq.25, which is justified because they vanish at LO
and are numerically suppressed at NLO for small values of $x$.
Therefore, Eq.25 at small values of $x$ can be rewritten as
\begin{eqnarray}
F_{L}^{c}(x,Q^{2})&=&C_{L,g}^{c}(x,Q^{2}){\otimes}G(x,Q^{2})\nonumber\\
&&{\equiv}\sum_{n=1}F^{(n),c}_{L}(x,Q^{2}),
\end{eqnarray}
where $n$ is the order of $\alpha_{s}$. Exploiting the derivatives
of the charm longitudinal structure function with respect to
${\ln}Q^{2}$ and inserting the DGLAP evolution equation, we find
that
\begin{eqnarray}
\frac{{\partial}F^{c}_{L}(x,Q^{2})}{{\partial}\ln
Q^{2}}&=&\frac{{d}{\ln}\alpha_{s}}{{d}\ln
Q^{2}}\sum_{n=1}[n{\times}F^{(n),c}_{L}(x,Q^{2})]\nonumber\\
&&+P_{gg}{\otimes}F^{c}_{L}(x,Q^{2})+\frac{d{\ln}C_{L,g}^{c}}{d{\ln
}Q^{2}}{\otimes}F^{c}_{L}(x,Q^{2}).\nonumber\\
\end{eqnarray}
%%%%%%%%%%%%%%%%%%%%%%%%%%%%%%%%%%%%%%%%%%%%%%%%%%%%%%%%%%%%%%%%%%%%%%
\subsection{III. Results and Discussions}
According to the last subsections we can determine gluonic
longitudinal structure function up to NNLO and also charm
longitudinal structure function up to NLO. The small-$x$ region of
the DIS offers a unique possibility to explore the Regge limit of
pQCD.  Phenomenologically, the Regge pole approach to DIS implies
that the charm structure function is sums of powers in $x$. The
simplest fit to the small-$x$ data corresponds to
$F^{c}_{L}(x,Q^{2})=f_{c}x^{-\lambda}$, where it is controlled by
pomeron exchange  small $x$. HERA shows that this behavior is
according to the gluon distribution small $x$, as
$g{\rightarrow}c\overline{c}$. In this limit, the gluon
distribution will become large, so its contribution to the
evolution of the parton distribution becomes dominant. Therefore
the gluon distribution has a rapid rise behavior  small $x$, that
is $xg(x,Q^{2})= f_{g}x^{-\lambda}$, where $\lambda$ is
corresponding to the hard-Pomeron intercept [30-33,62]. Exploiting
the small-$x$ asymptotic behavior for the gluon distribution and
charm structure functions to the evolution equations of the
gluonic longitudinal structure function and charm longitudinal
structure function respectively (Eqs.23, 27), evolution of the
longitudinal structure function at small-$x$ can be found as
\begin{eqnarray}
F_{L}(x,Q^{2})|_{x{\rightarrow}0}&=&F_{L}^{light}(x,Q^{2})({\rightarrow}F^{g}_{L}(x,Q^{2}))+F^{c}_{L}(x,Q^{2})\nonumber\\
&=&\sum_{n=1}F^{(n),g}_{L}(x,Q_{0}^{2})I^{(n)}_{g}\nonumber\\
&+&\sum_{n=1}F^{(n),c}_{L}(x,Q_{0}^{2})I^{(n)}_{c},
\end{eqnarray}
where
\begin{eqnarray}
I^{(n)}_{g}&=&\exp\Big(\int_{Q_{0}^{2}}^{Q^{2}}{d}\ln
Q^{2}(\sum_{n=1}n\frac{{d}{\ln}\alpha_{s}}{{d}\ln
Q^{2}}+P_{gg}{\otimes}x^{\lambda})\Big),\nonumber\\
\end{eqnarray}
and
\begin{eqnarray}
I^{(n)}_{c}&=&\exp\Big(\int_{Q_{0}^{2}}^{Q^{2}}{d}\ln
Q^{2}(\sum_{n=1}n\frac{{d}{\ln}\alpha_{s}}{{d}\ln
Q^{2}}+P_{gg}{\otimes}x^{\lambda}\nonumber\\
&&+\frac{{d}{\ln}C_{L,g}^{c}}{{d}\ln
Q^{2}}{\otimes}x^{\lambda})\Big).
\end{eqnarray}

Simplifying Eqs. (29) and (30) we get the compact forms

\begin{eqnarray}
I^{(n)}_{g}&=&
\prod_{n=1}\Big(\frac{{\alpha_{s}}_{n}(Q^2)}{{\alpha_{s}}_{n}(Q_{0}^2)}\Big)^{n}
{\cdot} \Big( \frac{Q^2}{Q_{0}^{2}}
\Big)^{P_{gg}{\otimes}x^{\lambda}},
\end{eqnarray}

and

\begin{eqnarray}
I^{(n)}_{c}&=&
\prod_{n=1}\Big(\frac{{\alpha_{s}}_{n}(Q^2)}{{\alpha_{s}}_{n}(Q_{0}^2)}\Big)^{n}
{\cdot} \Big(\frac{Q^2}{Q_{0}^{2}}
\Big)^{P_{gg}{\otimes}x^{\lambda}} \nonumber\\
&& {\cdot} \Big( \frac{C_{L,g}^{c}(Q^2)}{C_{L,g}^{c}(Q_{0}^2)}
\Big)
 {\otimes} x^{\lambda}.
\end{eqnarray}

In Figs.1-3, we present the small $x$ behavior of the $F_{L}$
  structure function according to the evolution equation (28) as a function of
  $x$  for $Q^{2}=12$, $45$ and $120~ GeV^{2}$. In the left
  hand of these figures, we present the heavy contribution $F_{L}^{c}$, gluonic contribution
   $F_{L}^{g}$ and total $F_{L}^{Total}$ (heavy + gluonic) of longitudinal structure function
    with results of DL [30-33,62] and CD models [63]. In the right hand side, $F_{L}^{Total}$
     has been presented with H1 data [54] with total error and on-shell limit of the $k_{T}$
factorization [64], where the transverse momentum of the gluon
$k^{2}$ is much smaller than the virtuality of the photon (
$k^{2}{<<}Q^{2}$) and this is consistent with the collinear
factorization as the $k_{T}$ factorization formula can be
determined from the inclusive cross section in dipole
representation.
      In all the cases longitudinal structure function $F_L$ increases towards smaller $x$
      for a fixed $Q^2$. We compared our $F_{L}^{g}$ results with the results of DL model
      and $F_{L}^{c}$ results with the results of CD model. We observed that our $F_{L}^{c}$ results are somewhat higher than those of CD model in all $Q^2$. But though our  $F_{L}^{g}$ results are somewhat higher than those of DL model, their differences decreases when  $Q^2$ increase and they almost coincide at $120 GeV^2$. On the otherhand, we observed that H1 data have been well described by our results as well as the results of $k_{T}$ factorization. Of course our results are slightly above than those of $k_{T}$ factorization. When $Q^2$ increases, our results become in better agreement with the data.\\
In Figs. 4-5, we present the same results of $F_{L}$
  structure function as a function of  $Q^2$ for small-$x$ values $x=0.001$ and $x=0.0004$.
  In all the cases longitudinal structure function $F_L$ increases towards higher $Q^2$ for
  a fixed $x$ and smaller $x$ for a fixed $Q^2$. Our $F_{L}^{g}$ results are very close to
   DL results especially at $x=0.001$. But our  $F_{L}^{c}$ results are slightly higher
    than those of CD results. Also it is observed that the rate of increment of heavy
    longitudinal structure function $F_{L}^{c}$ is more than that of gluonic longitudinal
    structure function  $F_{L}^{g}$, and both approach to closer values towards higher $Q^2$
    values. Again comparing the $Q^2$-evolutions of total (heavy + gluonic) longitudinal
    structure function $F_{L}^{Total}$ with the results of $k_{T}$
factorization, it is seen
    that our results are comparable to H1 results, especially at higher-$x$, $x=0.001$; but somewhat higher in smaller-$x$, $x=0.0004$ in higher $Q^2$. On the other hand, $k_{T}$ factorization results are better in all the cases. \\
 In our calculations, we use
the DL model of the gluon distribution and also we set
the running coupling constant according to the Table 1. For heavy
contribution to the longitudinal structure function,  we choose
$m_{c}=1.3~GeV$ and the renormalization scale is
$<\mu^{2}>=4m^{2}_{c}+Q^{2}/2$.\\

\subsection{IV. Conclusions}

In conclusion, we have observed that the hard-pomeron behaviour for the longitudinal structure function
dynamical behaviour gives the heavy quark effects to the light flavours at small-$x$. We can see in all figures the increase of our results
for longitudinal structure function $F_{L}(x,Q^{2})$ towards
smaller $x$  and higher $Q^2$ which is consistent with QCD
calculations, reflecting the rise of gluon and charm (heavy)
distribution inside proton in this region. Our results for gluonic
and charm (heavy) longitudinal structure function do not exactly
tally with results of DL and CD models respectively, as formers
are somewhat higher than laters. Though $F_{L}^{g}$ is more or less
comparable with results of DL models, $F_{L}^{c}$ is somewhat
higher than that of CD models results. Our total results $F_{L}^{Total}$ and those of  $k_{T}$ factorization are well within the data range.
 Lastly, one important conclusion is that charm (heavy) contribution to total longitudinal
 structure function is considerable one and can not be neglected especially at smaller $x$ and higher $Q^2$ region.

%%%%%%%%%%%%%%%%%%%%%%%%%%%%%%%%%%%%%%%%%%%%%%%%%%%%%%%%%%%%%%%%%%%%%%%%%
\subsection{Appendix A}
The explicit forms of the first-, second- and third-order
splitting functions are respectively [49-52]
\begin{eqnarray}
P^{\rm LO}_{gg}(z)&=&2C_{A}(\frac{z}{(1-z)_{+}}+\frac{(1-z)}{z}+z(1-z))\nonumber\\
&&+\delta(1-z)\frac{(11C_{A}-4N_{f}T_{R})}{6},
\end{eqnarray}
where the $^{,}\text{plus}^{,}$ distribution is defined by
\begin{equation}
\int_{0}^{1} dz \frac{f(z)}{(1-z)_{+}}=\int_{0}^{1}dz
\frac{f(z)-f(1)}{1-z},\nonumber
\end{equation}

\begin{eqnarray}
P_{gg}^{\rm NLO}&=&C_FT_F(-16+8z+\frac{20}{3}z^2+\frac{4}{3z}\nonumber\\
&&-(6+10z)\ln z-(2+2z)\ln ^{2}z)\nonumber\\
&&+C_AT_F(2-2z+\frac{26}{9}(z^2-z^{-1})\nonumber\\
&&-\frac{4}{3}(1+z)\ln z-\frac{20}{9}P_{gg}(z))\nonumber\\
&&+C_A^2(\frac{27}{2}(1-z)+\frac{67}{9}(z^2-z^{-1})\nonumber\\
&&-(\frac{25}{3}-\frac{11}{3}z+\frac{44}{3}z^2)\ln z\nonumber\\
&&+4(1+z)\ln ^{2}z+2P_{gg}(-z)S_2(z)\nonumber\\
&&+(\frac{67}{9}-4\ln z\ln(1-z)\nonumber\\
&&+\ln ^{2}z-\frac{\pi^2}{3})P_{gg}(z)),\nonumber\\
\end{eqnarray}

where
\begin{eqnarray}
P_{gg}(z)=\frac{1}{(1-z)_{+}}+\frac{1}{z}-2+z(1-z),\nonumber
\end{eqnarray}
and the function $
S_2(z)=\int_{\frac{z}{1+z}}^{\frac{1}{1+z}}\frac{dy}{y}{\ln}(\frac{1-y}{y})$
is defined in terms of the dilogarithm function  as
\begin{eqnarray}
S_{2}(z)=-2Li_{2}(-z)+\frac{1}{2}\ln ^{2}z-2\ln
z\ln(1+z)-\frac{\pi^{2}}{6},\nonumber
\end{eqnarray}

and

\begin{eqnarray}
P_{gg}^{\rm NNLO}&=&2643.521D0+4425.894\delta(1-z)\nonumber\\
&&+3589L1-20852+3968z-3363z^2\nonumber\\
&&+4848z^3+L0L1(7305+8757L0)\nonumber\\
&&+274.4L0-7471L0^2+72L0^3-144L0^4\nonumber\\
&&+14214z^{-1}+2675.8z^{-1}L0\nonumber\\
&&+N_f(-412.172D0-528.723\delta(1-z)\nonumber\\
&&-320L1-350.2+755.7z-713.8z^2\nonumber\\
&&+559.3z^3+L0L1(26.15-808.7L0)\nonumber\\
&&+1541L0+491.3L0^2+\frac{832}{9}L0^3\nonumber\\
&&+\frac{512}{27}L0^4+182.96z^{-1}\nonumber\\
&&+157.27z^{-1}L0)+N_f^2(-\frac{16}{9}D0\nonumber\\
&&+6.4630\delta(1-z)-13.878+153.4z\nonumber\\
&&-187.7z^2+52.75z^3\nonumber\\
&&-L0L1(115.6-85.25z+63.23L0)\nonumber\\
&&-3.422L0+9.680L0^2-\frac{32}{27}L0^3\nonumber\\
&&-\frac{680}{243}z^{-1}),
\end{eqnarray}

where $L0=\ln z,~~ L1=\ln(1-z)$ and $D0=\frac{1}{(1-z)_{+}}$.\\
%%%%%%%%%%%%%%%%%%%%%%%%%%%%%%%%%%%%%%%%%%%%%%%%%%%%%%%%%

\subsection{Appendix B}
The $C^{c}_{L,g}$ is the charm coefficient longitudinal function
in LO and NLO analysis and it is given by
\begin{eqnarray}
C_{L,g}(z,\zeta)&{\rightarrow}&C^{(0)}_{L,g}(z,\zeta)+a_{s}(\mu^{2})[C_{L,g}^{(1)}(z,\zeta)\\\nonumber
&&+\overline{C}_{L,g}^{(1)}(z,\zeta)ln\frac{\mu^{2}}{m_{c}^{2}}].
\end{eqnarray}
In LO analysis, this coefficient can be found in Refs.[15-16], as
\begin{eqnarray}
C^{(0)}_{g,L}(z,\zeta)=-4z^{2}{\zeta}ln\frac{1+\beta}{1-\beta}+2{\beta}z(1-z),
\end{eqnarray}
where $\beta^{2}=1-\frac{4z\zeta}{1-z}$ and $\mu$ is the mass
factorization scale, which has been put equal to the
renormalization scales $\mu^{2}=4m_{c}^{2}$ or
$\mu^{2}=4m_{c}^{2}+Q^{2}$, and in the NLO analysis we can use the
compact form of these coefficients according to the
Refs.[17-21].\\
%%%%%%%%%%%%%%%%%%%%%%%%%%%%%%%%%%%%%%%%%%%%%%%%%%%%%%%%%%%%%%%%%%%%%%%%
%%%%%%%%%%%%%%%%%%%%%%%%%%%%%%%%%%%%%%%%%%%%%%%%%%%%%%%%%%%%%%%%
\begin{table}
\centering \caption{The QCD coupling and corresponding $\Lambda$
parameter for $N_{f}=4$, for LO, NLO and NNLO fits according to
Refs.[65-67].}\label{table:table1}
\begin{minipage}{\linewidth}
\renewcommand{\thefootnote}{\thempfootnote}
\centering
\begin{tabular}{|l|c|c|} \hline\noalign{\smallskip}  & $ \alpha_{s}(M_{Z}^{2})$ &$
\Lambda_{QCD}(MeV)$  \\
\hline\noalign{\smallskip}
LO & 0.130 & 220 \\
NLO & 0.119 & 323 \\
NNLO & 0.1155 & 235 \\
\hline\noalign{\smallskip}
\end{tabular}
\end{minipage}
\end{table}
%%%%%%%%%%%%%%%%%%%%%%%%%%%%%%%%%%%%%%%%%%%%%%%%%%%%%%%%%%%%

\textbf{References}\\
1. A. Gonzalez-Arroyo, C. Lopez, and F.J. Yndurain, Phys. Lett. B98- 218(1981).\\
2. A. M. Cooper-Sarkar, G. Inglman, K. R. Long, R. G. Roberts, and D. H. Saxon , Z. Phys. C39- 281(1988).\\
3. R. G. Roberts, The structure of the proton, (Cambridge University Press 1990)Cambridge.\\
4. Aaron, F. D., et al., (H1 Collaboration), Phys.
Lett. B665- 139(2008).\\
5. A. D. Martin, W. J. Stirling, R. S. Thorne, G. Watt, Eur. Phys. J. C70- 51(2010).\\
6. S. Moch and A.Vogt, JHEP0904- 218(1981).\\
7. R. S. Thorne, Phys. Lett.B418- 371(1998).\\
8. R. S. Thorne, arXiv:hep-ph/0511351 (2005).\\
9. A. D. Martin, W. J. Stirling, R. S.Thorne, Phys. Lett. B
635- 305(2006).\\
10. A. D. Martin, W. J. Stirling, R. S.Thorne, Phys. Lett. B636- 259(2006).\\
11. S. Moch, J. Vermaseren and A. Vogt, Nucl. Phys. B688-
101(2004).\\
12. S. Moch, J. Vermaseren and A. Vogt, Nucl. Phys. B691-
129(2004).\\
13. Moch, J. Vermaseren and A. Vogt, Phys. Lett. B606- 123(2005).\\
14. M. Gluck, C. Pisano and E. Reya, Phys. Rev. D77- 074002 (2008).\\
15. M. Gluk, E. Reya and A. Vogt, Z. Phys. C67- 433(1995).\\
16. M. Gluk, E. Reya and A. Vogt, Eur. Phys. J. C5- 461(1998).\\
17. E. Laenen, S. Riemersma, J. Smith and W. L. van Neerven,
Nucl. Phys. B392- 162(1993).\\
18. A.~Y.~Illarionov, B.~A.~Kniehl and A.~V.~Kotikov, Phys.
Lett.B663- 66(2008).\\
19. S. Catani, M. Ciafaloni and F. Hautmann, Preprint
CERN-Th.6398/92, in Proceeding of the Workshop on Physics at HERA
(Hamburg)2-690(1991).\\
20. S. Catani and F. Hautmann, Nucl. Phys. B427- 475(1994).\\
21. S. Riemersma, J. Smith and W. L.
van Neerven, Phys. Lett. B347- 143(1995).\\
22. E. A. Kuraev, L. N. Lipatov and V. S. Fadin, Sov.Phys.JETP44- 443(1976).\\
23. E. A. Kuraev, L. N. Lipatov and V. S. Fadin, Sov. Phys. JETP45- 199(1977).\\
24. Y. Y. Balitsky and L. N. Lipatov, Sov. Journ. Nucl. Phys.28- 822(1978).\\
25. L.V.Gribov, E.M.Levin and M.G.Ryskin, Phys.Rep.100- 1(1983).\\
26. J. Kwiecinski, arXiv:hep-ph/9607221 (1996).\\
27. J. Kwiecinski, A.D.Martin and P.J.Sutton, Phys.Rev.D44- 2640(1991).\\
28. K.Kutak and A.M.Stasto, Eur.Phys.J.C41-
 343(2005).\\
29. S.Bondarenko, arXiv:hep-ph/0808.3175(2008).\\
30. A. Donnachie and  P. V. Landshoff, Phys. Lett. B296- 257(1992).\\
31. A. Donnachie and P. V. Landshoff, Phys. Lett. B437- 408(1998 ).\\
32. A. Donnachie and P. V. Landshoff, Phys. Lett. B550- 160(2002 ).\\
33. P.V.Landshoff, arXiv:hep-ph/0203084(2002).\\
34. P. Desgrolard, M. Giffon, E. Martynov and E. Predazzi, Eur. Phys. J. C18- 555(2001).\\
35. P. Desgrolard, M. Giffon and E. Martynov, Eur. Phys. J. C7- 655(1999).\\
36. A.A.Godizov, Nucl.Phys.A927 36(2014).\\
37. Yu. L. Dokshitzer, Sov. Phys. JETP6- 641(1977 ).\\
38.  G. Altarelli and
G. Parisi, Nucl. Phys. B126- 298(1997 ).\\
39. V. N. Gribov and L. N. Lipatov, Sov. J. Nucl. Phys.28- 822(1978).\\
40. M. B. Gay Ducati  and V. P. B. Goncalves, Phys. Lett. B390- 401(1997).\\
41. K. Pretz, Phys.Lett.B311- 286(1993).\\
42. K. Pretz, Phys. Lett. B332- 393(1994).\\
43. A. V. Kotikov, arXiv:hep-ph/9507320(1995).\\
44. G. R. Boroun, JETP, Vol. 106,4- 701(2008).\\
45. G. R. Boroun and B. Rezaei, Eur. Phys. J. C72- 2221(2012).\\
46.  G. R. Boroun and B. Rezaei, Eur. Phys. J. C73- 2412(2013).\\
47.  M. Devee, R Baishya and J. K. Sarma, Eur. Phys.J. C72- 2036(2012).\\
48.  R. Baishya, U. Jamil and J. K. Sarma, Phys. Rev. D79- 034030(2009).\\
49. S. Moch and A. Vogt, JHEP0904- 081(2009).\\
50. R. K. Ellis , W. J. Stirling and B. R. Webber, QCD and
Collider Physics(Cambridge University Press)(1996).\\
51. C. Pisano, arXiv:hep-ph/0810.2215 (2008).\\
52. A. Retey, J. Vermaseren , Nucl. Phys. B604- 281(2001).\\
53. B. Lampe, E. Reya, Phys. Rep.332- 1(2000).\\
54. V. Andreev et al. [H1
Collaboration], Accepted by Eur. Phys. J. C, arXiv:hep-ex/1312.4821 (2013).\\
55. N. Ya. Ivanov, Nucl. Phys. B814- 142(2009).\\
56.  N. Ya. Ivanov, Eur. Phys. J. C59- 647(2009).\\
57. I. P. Ivanov and N. Nikolaev, Phys. Rev. D65- 054004(2002).\\
58. G. R. Boroun and B. Rezaei, Nucl. Phys. B857-143(2012).\\
59. G. R. Boroun and B. Rezaei, Eur. Phys. Lett100- 41001(2012).\\
60. G. R. Boroun and B. Rezaei, Nucl. Phys. A929- 119(2014).\\
61. G.R.Boroun, Nucl. Phys. B884- 684(2014).\\
62. R. D. Ball and P. V. landshoff, J. Phys. G26- 672(2000).\\
63. N. N. Nikolaev and V. R. Zoller, Phys. Lett. B509-
283(2001).\\
64. K. Golec-Biernat and A. M. Stasto, Phys. Rev. D80
014006(2009).\\
65. A. D. Martin, R. G. Roberts, W. J. Stirling, R. S. Thorne, Phys. Lett. B531- 216(2002).\\
66. A. D. Martin, R. G. Roberts, W. J. Stirling, R. S. Thorne, Eur. Phys. J. C23- 73(2002).\\
67. A. D. Martin, R. G. Roberts, W. J. Stirling, R. S. Thorne, Phys. Lett. B604- 61(2004).\\

%%%%%%%%%%%%%%%%%%%%%%%%%%%%%%%%%%%%%%%%%%%%%%
%\small\textbf{Acknowledgment}\\

%Our sincere gratitude goes to Razi University for its financial
%support in carrying out this study.\\
%%%%%%%%%%%%%%%%%%%%%%%%%%%%%%%%%%%%%%%%%%%%%
\begin{figure}
\includegraphics[width=0.5\textwidth]{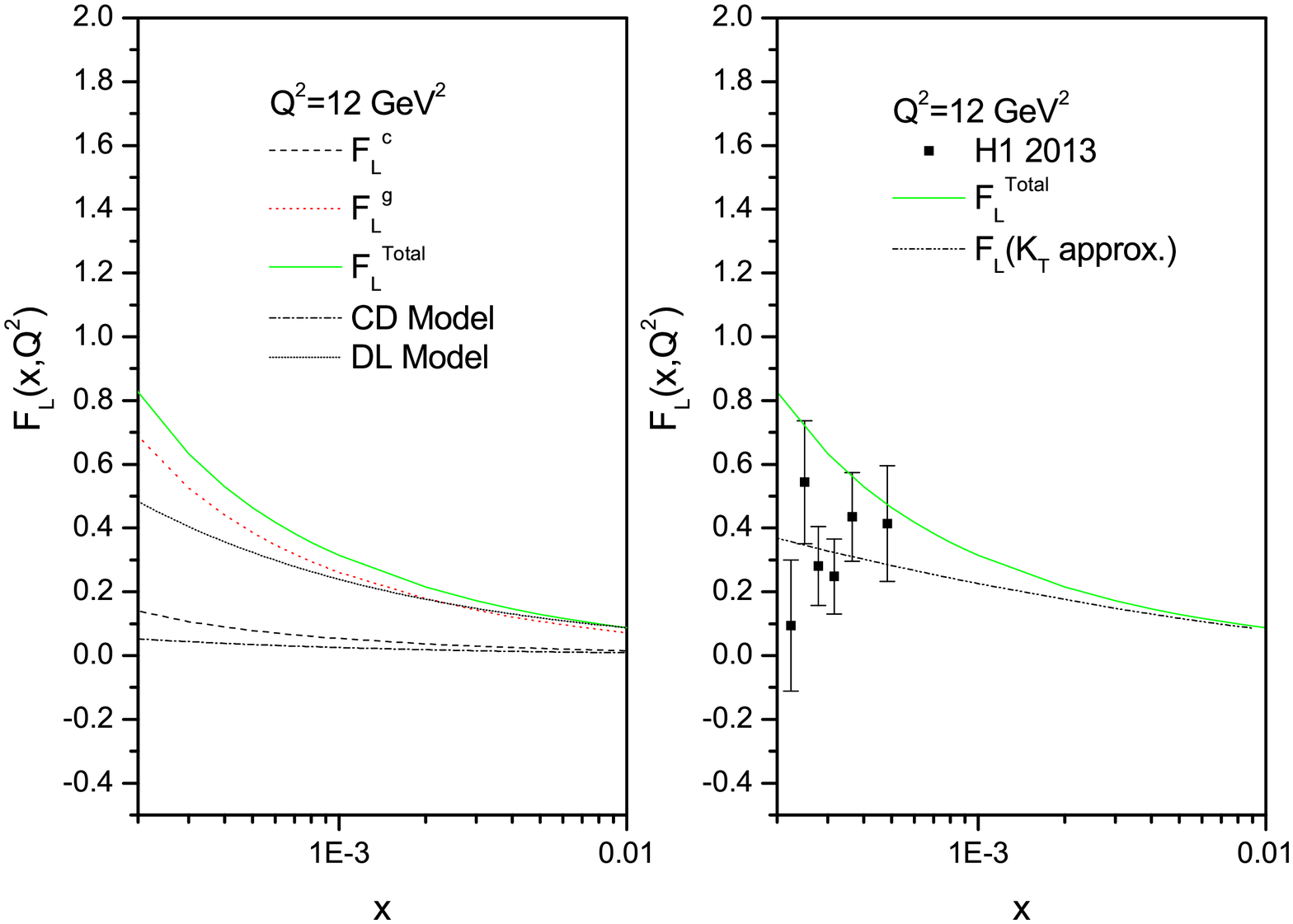}
\caption{{\textit{left}}: Dynamical light and heavy contributions
to the total $F_{L}$  small $x$ for $Q^{2}=12~GeV^{2}$ at NNLO
analysis, compared with DL [30-33,62] and CD [63] models respectively.\\
{\textit{right}}: The total $F_{L}$ compared with $k_{T}$
factorization [64] and H1 data [54] with total error.}\label{Fig1}
\end{figure}
\begin{figure}
\includegraphics[width=0.5\textwidth]{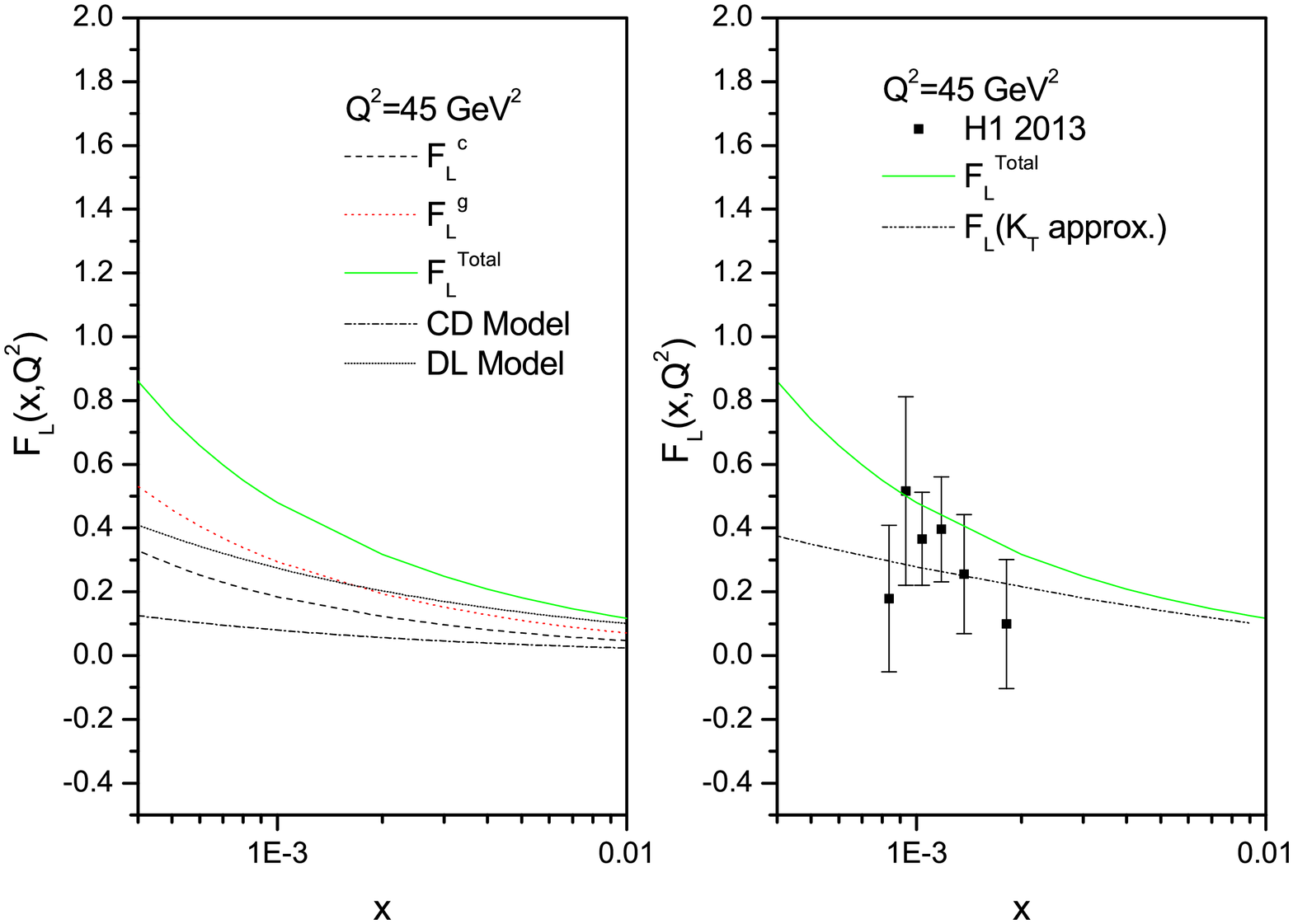}
\caption{As in Fig. 1 but for $Q^{2}=45~GeV^{2}$. }\label{Fig2}
\end{figure}
\begin{figure}
\includegraphics[width=0.5\textwidth]{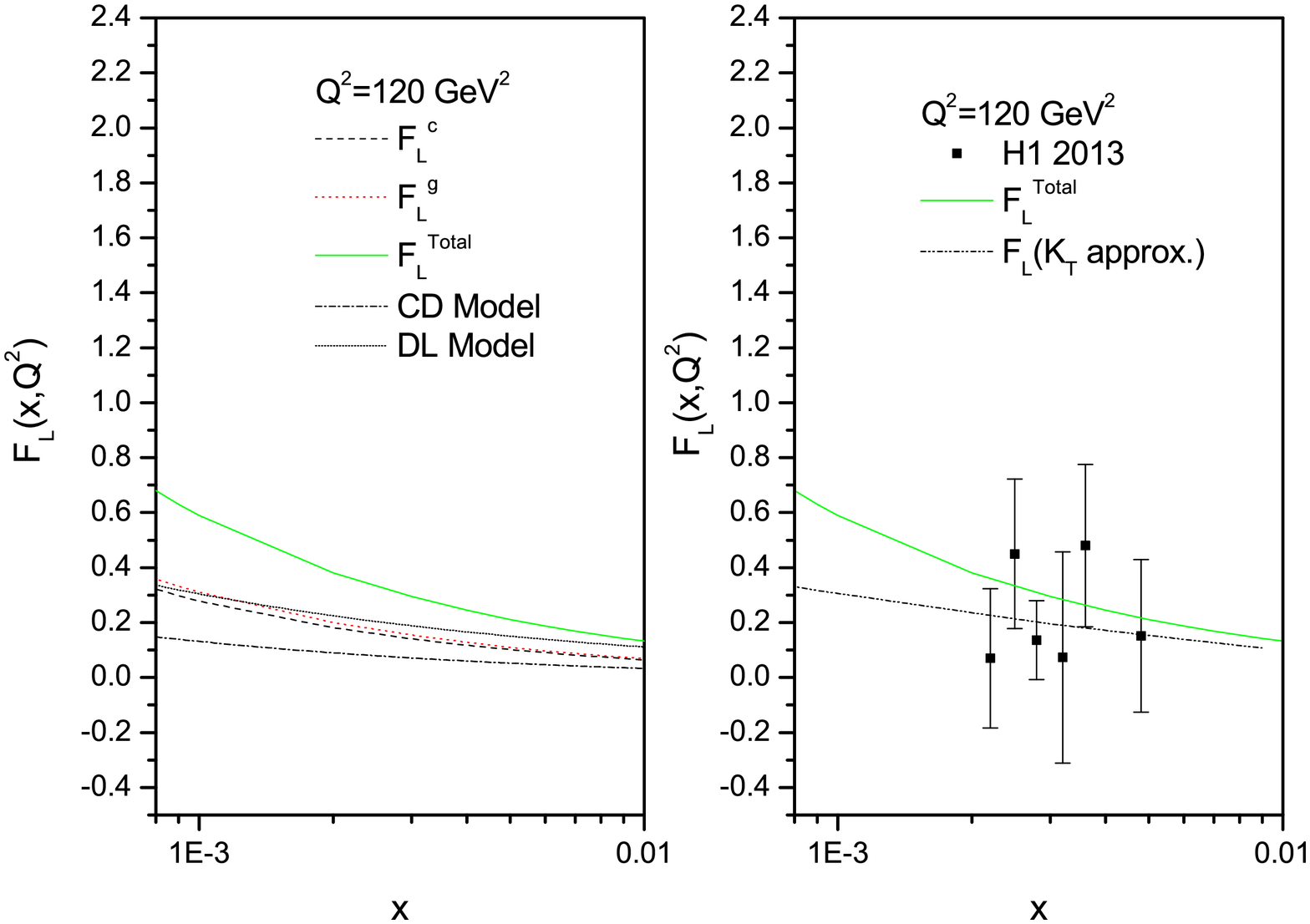}
\caption{As in Fig. 1 but for $Q^{2}=120~GeV^{2}$. }\label{Fig3}
\end{figure}

\begin{figure}
\includegraphics[width=0.5\textwidth]{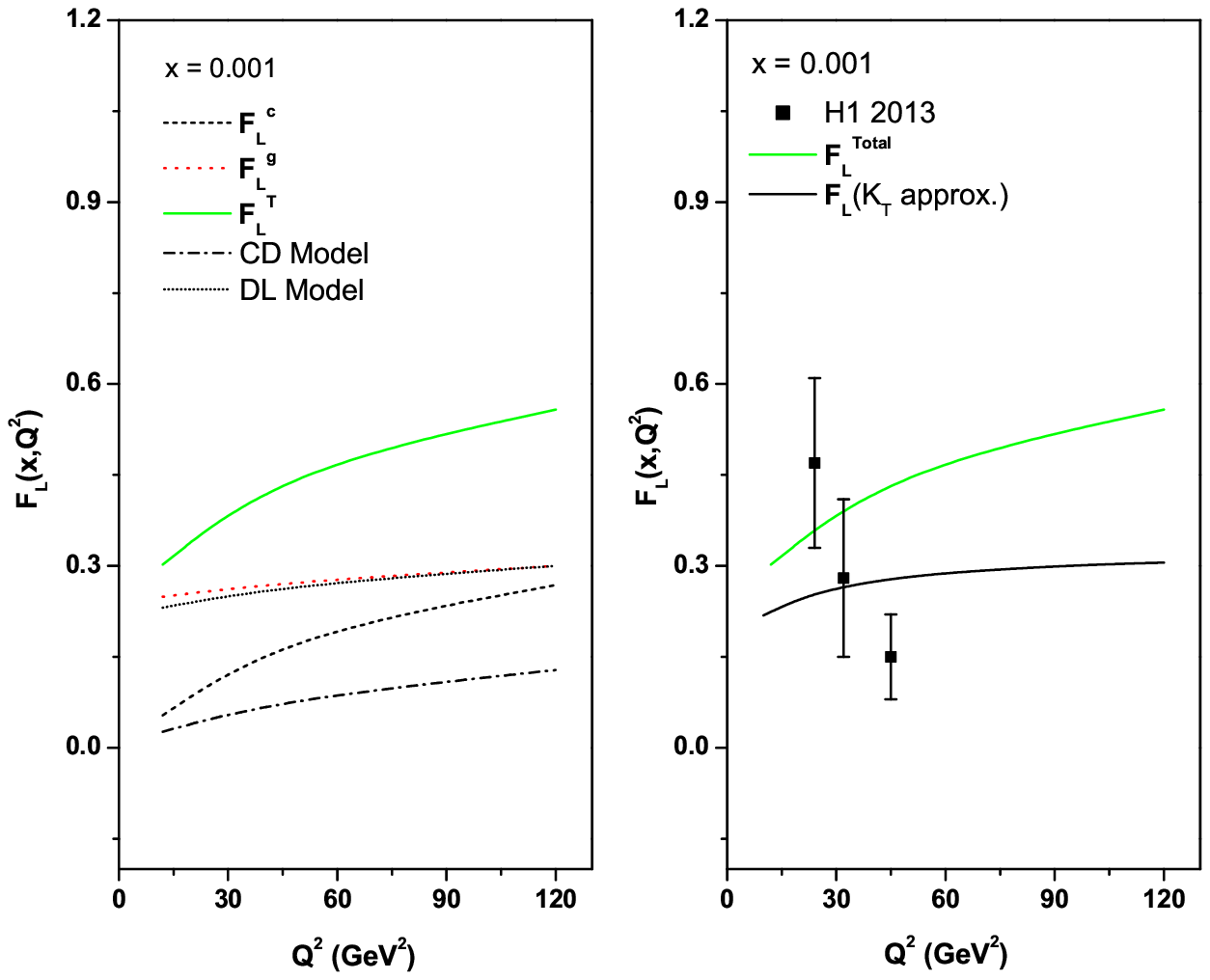}
\caption{{\textit{left}}: Dynamical light and heavy contributions
to the total $F_{L}$  small $x$ for $x=0.001$ at NNLO
analysis, compared with DL [30-33,62] and CD [63] models respectively.\\
{\textit{right}}: The total $F_{L}$ compared with $k_{T}$
factorization [64] and H1 data [54]  with total
error.}\label{Fig4}
\end{figure}
\begin{figure}
\includegraphics[width=0.5\textwidth]{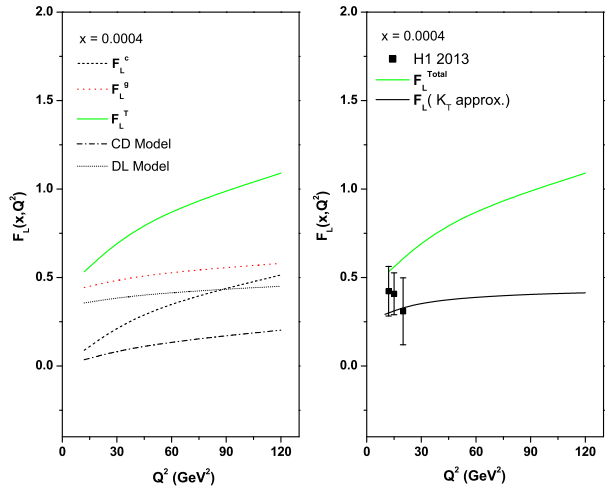}
\caption{As in Fig. 4 but for $x=0.0004$. }\label{Fig5}
\end{figure}

%%%%%%%%%%%%%%%%%%%%%%%%%%%%%%%%%%%%%%%%%%%%%%%%%%%%%%%%%%%%
\end{document}